# Tunable band gap in germanene by surface adsorption

Meng Ye,[1,†] Ruge Quhe,[1,3,†] Jiaxin Zheng,[1,2] Zeyuan Ni,[1] Yangyang Wang,[1] Yakun Yuan,[1] Geoffrey Tse,[1] Junjie Shi,[1] Zhengxiang Gao,[1] and Jing Lu[1,2*]

[1] State Key Laboratory of Mesoscopic Physics and Department of Physics,

Peking University, Beijing 100871, P. R. China

[2] Collaborative Innovation Center of Quantum Matter, Beijing 100871, China

[3] Academy for Advanced Interdisciplinary Studies,

Peking University, Beijing 100871, P. R. China

†These authors contributed equally to this work.

*Corresponding author: jinglu@pku.edu.cn

## Abstract

Opening a sizable band gap in the zero-gap germanene without heavy loss of carrier mobility is a key issue for its application in nanoelectronic devices such as high-performance field effect transistors (FETs) operating at room temperature. Using the first-principles calculations, we find a band gap is opened at the Dirac point in germanene by single-side adsorption of alkali metal (AM) atoms. This band gap is tunable by varying the coverage and the species of AM atoms, ranging from 0.02 to 0.31 eV, and the maximum global band gap is 0.26 eV. Since the effective masses of electrons and holes in germanene near the Dirac point after surface adsorption (ranging from 0.005 to 0.106 $m_e$) are small, the carrier mobility is expected not to degrade much. Therefore germanene is a potential candidate of effective FET channel operating at room temperature upon surface adsorption.

**Introduction**

Silicene and germanene are silicon and germanium analogies of graphene. Theoretical researches have predicted that they prefer low-buckled (LB) structures to the planar ones[1-3]. And they have electronic structures which are very similar to that of graphene: they are zero-gap semi-metals and their charge carriers are massless Dirac fermions[1,2,4,5]. So they are expected to have similar properties with graphene, especially the high mobility of charge carriers. In experiment, there are several reports about the synthesis of pristine silicene on Ag/Ir (111) and $ZrB_2$ surface[6-13]. Besides, multilayer hydrogen-terminated germanene has also been synthesized by using the topochemical deintercalation of $CaGe_2$ very recently, and it can be mechanically exfoliated to single layer onto $SiO_2$/Si surface[14].

However, the zero-gap properties of silicene and germanene make them very difficult to deploy in nature, when applying in electronic devices such as high-performance field effect transistors (FETs) operating at room temperature. Therefore, opening a sizable band gap in them without heavy loss of carrier mobility is one of the key issues in present research on them. Several methods have been developed to do the job. For instance, Ni *et al.* reported that when an vertical electric field is applied to silicene/germanene, the atoms in LB structure are no longer equivalent, and a tunable band gap is opened. The band gap increases linearly with the electric field, and the maximum band gap under an experimentally accessible electric field is only 0.03 eV[15]. Besides, two gates are required in fabrication of a silicene/germanene based FET in order to create a band gap and tune channel's conductance individually.

Single-side adsorption is an alternative method to open a band gap in silicene[16]. The mechanism of this method to open a band gap is the sublattice/bond symmetry breaking which induced by the charge transfer from alkali metal (AM) atoms to silicene. The band gap is tunable by varying the coverage of AM, and the maximum band gap of silicene by AM atoms adsorption is 0.5 eV, which meets the requirement of a logic device. Besides the larger band gap, the FETs made of silicene adsorbed by AM atoms need only one gate. Such method applied will make the device operating in a more effective way and will possibly advance over the previous one.

In this article, we investigate germanene with AM (lithium, sodium, and potassium) atoms adsorption by using the density functional theory (DFT) calculations. As we predicted, a band gap is opened at the Dirac point and it is tunable by controlling the species and coverage of AM. The maximum global band gap we obtain is 0.26 eV. Since the effective mass of electrons and holes near Dirac point are small, we expected the carrier mobility won't loss heavily after adsorption. These facts suggest the potential of germanene with AM atoms adsorption to fabricate electronic device such as high-performance FETs.

## Computational Details

Geometry optimization and electronic structure of all structures in this article are calculated by using all-electron double numerical atomic basis set plus polarization (DNP), as implemented in the DMol$^3$ package[17,18]. The generalized gradient approximation (GGA) to exchange-correlation functional of the Perdew-Wang 91 (PW91) form[19] is used with inclusion of semi-empirical dispersion correction of Ortmann-Bechstedt-Schmidt (OBS) form[20] (DFT-D) in this article. We place a 40 Å of vacuum layer between the periodic layers to avoid any possible interactions. A Monkhorst-Pack $k$-points grid[21] with 0.01 Å$^{-1}$ separations (which means $28 \times 28 \times 1$ grid for pristine germanene) is used in the first Brillouin zone sampling. The geometry optimization is carried out until the max force on each atom is less than $2.7 \times 10^{-4}$ eV/Å. We set the orbital cutoff to 5.6 Å in order to obtain the correct adsorption energy when we performed the energy calculations. The component and wave function of the energy band are analyzed with resort to additional calculations based on the plane-wave basis set with a cut-off energy of 380 eV and the projector-augmented wave (PAW) pseudopotential, implemented in the Vienna *ab initio* simulation package (VASP).[22-24] The band structures generated by these two packages are very close. We also use VASP code to investigate the spin-orbit coupling (SOC) effects in our materials.

As shown in Figure 1, we use the $m \times m$ supercell model, which consists of a $m \times m$ ($m = 1, \sqrt{3}, 2, 3, 2\sqrt{3}, 4$) germanene primitive cell with one AM atom on the top, to investigate the germanene with AM atoms adsorption. When $m = 2, 2\sqrt{3}, 4$, we also put two AM atoms on the top, which gives the $m \times m/2$ model (since potassium atom has quite large radius, KGe$_2$ and KGe$_4$ are physically impossible). The structures with different coverage of AM atoms are denoted as AMGe$_n$, where $n$ is the stoichiometric ratio of germanium to AM per supercell). Here we define the coverage of AM $N = 1/n$. we can simply derive that $N = 1/2m^2$ and $1/m^2$, for the $m \times m$ and $m \times m/2$ supercell model respectively.

## Results and Discussion

**Geometry and stability**

We consider three adsorption configurations of the AM atoms: above the center of a hexagonal ring and on the top of two inequivalent germanium atoms. The first type of configuration (as shown in Figure 1) is the most stable. Several previous works also suggest that AM atoms-covered graphene/silicene has the configuration where the AM atoms are located above the center of a hexagonal ring[16,25]. These facts provide strong support for our results.

The optimized structure of germanene has an in-plane lattice constant $a = 4.063$ Å and a buckling $\Delta = 0.677$ Å. These results are in good agreement with previous works[2-5,15]. Since there are more than one buckling values in germanene after the adsorption, we only give the maximum one in each structure here. This value increases with the increasing coverage, ranging from 0.803 ~ 1.208, 0.775 ~ 1.176, 0.754 ~ 0.869 Å for Li, Na, and K adsorption, respectively. The distance from the AM atoms to the nearest germanium atoms ($d_1$, as shown in Figure 1a) is almost independent of the coverage, but strongly depends on the species of AM. As the radius of AM atoms increases, $d_1$ increases from 2.7 to 3.4 Å.

To measure the stability of germanene covered by AM atoms, here we define the adsorption energy ($E_a$) as

$$E_a = E_{Ge} + E_{AM} - E_{Ge+AM} \quad (1)$$

where $E_{Ge}$, $E_{AM}$, and $E_{Ge+AM}$ are the relaxed energy for germanene, isolated AM atom, and combined system, respectively. As Figure 2 shows, $E_a$ of all the investigated systems are 1.65 to 2.38 eV per AM atom, which suggests that the adsorption of AM atoms on germanene is rather strong. The adsorption energy of KGe$_n$ decrease as the coverage increases, but those of LiGe$_n$ and NaGe$_n$ first decrease and then increase with the increasing coverage. We can understand the decreasing trend of $E_a$ with the increasing coverage, under the condition of low coverage in terms of two facts. Firstly, the charge transfer to germanene per AM atom decreases with the increasing coverage (as shown in Figure 3b), so the ionic attraction between the AM atoms and germanene also becomes weaker as the coverage increases. Second, a larger coverage would bring the AM atoms closer, resulting in a stronger ionic repulsion between them. With further increase of the coverage, the gradual formation of metallic bond between AM atoms makes the adsorption energy increase. Besides, we notice that although the transferred charge from Li atoms to germanene is much smaller than those of Na and K atoms at the same coverage, the adsorption energy of LiGe$_n$ is larger than those of NaGe$_n$ and KGe$_n$. Therefore, a strong Li-Ge coverlent interaction exists in LiGe$_n$ besides the ionic attraction.

**Electronic structure**

Figure 4a shows the band structure of germanene, which is quite similar to those of graphene and silicene. The energy dispersion is linear near the Dirac point $K$. The charge carries of germanene, electrons and holes, behave like massless Dirac fermions. Here we use $v_F$, the Fermi velocity of charge carriers, to represent the dispersion of the $\pi^*$ and $\pi$ bands near the point $K$

$$E(\boldsymbol{K} + \boldsymbol{k}) = E_F \pm \hbar v_F k \qquad (2)$$

where $E_F$ is the Fermi energy. We calculate the Fermi velocity of charge carriers in germanene by fitting the dispersion of the $\pi^*$ and $\pi$ band near the Dirac point, using Eq. (2). The result we obtain is $(5.2 \pm 0.1) \times 10^5$ m/s, which is in good agreement with the previous work[26].

The electronic structures of Li-adsorbed germanene with different coverages are shown in the left panel of Figure 4b-4d and Figure S1 of Supplement Information. The Dirac cone is basically maintained in LiGe$_2$, and a band gap of $E_g = 0.13$ eV is opened between the $\pi$ and $\pi^*$ band at the Dirac point. This band gap becomes 0.08/0.26 eV in LiGe$_4$/LiGe$_6$. In fact, all the investigated LiGe$_n$ have a band gap at the Dirac point, a band gap of about 0.02 eV can still be observed even when the coverage is rather small ($N = 1/32$), as shown in Figure S1f. Germanene with sodium and potassium atoms adsorption have similar band structures with lithium atoms-covered germanene at the same coverage, but with a different value of band gap opened at the Dirac point. The band gap as a function of coverage is shown in Figure 5a, and it shows an increase tendency with the increasing coverage, with especial increase at $n = 3x$, generally.

We then will discuss the mechanism of band gap opening in germanene after the adsorption. Obviously, the change transfer between AM atoms and germanene gives rise to a built-in electric field (as shown in Figure 3a), which breaks the sublattice symmetry in germanene and opens a band gap at the Dirac point. We use a parallel plate capacitor model to estimate the strength of the built-in electric field ($E$), which is calculated by using Eq. (3)

$$E = \frac{2Q_{Ge}}{\varepsilon_0 a^2 \sin\frac{\pi}{3}} \tag{3}$$

where $Q_{Ge}$, $\varepsilon_0$, and $a$ are the Mulliken transferred charge density per germanium atom, the vacuum permittivity, and the lattice constant of germanene, respectively. Previous DFT calculations have shown that the band gap of germanene increases linearly with the strength of the external vertical electric field, with a slop of about 0.1 e · Å[15]. The solid black line in Figure 5b represents the ideal $E_g$-$Q_{Ge}$ relation generated by Eq. (3). We find that the actual band gaps are distributed around this straight line, with a good agreement in small band gap cases. Therefore, the band gap in germanene opened by AM atoms adsorption is chiefly attributed to built-in field induced by the transferred charge. Because the strength of built-in electric field can be as large as 3 V/Å (in NaGe$_2$), which is one order of magnitude larger than that of the largest experimentally accessible external electric field (about 0.6 V/Å[27]), our method can open a band gap 10 times larger than that obtained by applying an external electric field (0.03 eV).

However, the charge transfer and sublattice symmetry breaking mechanism cannot explain all the results, since AMGe$_{3x}$ has a notably larger band gap and AMGe$_2$/AMGe$_4$ has a smaller one. A previous work has suggested that in Li-adsorbed graphene, only LiC$_{3x}$ have a band gap at the Dirac point[28]. Since graphene prefers planer structure, the transferred charge from Li atoms to graphene cannot break the sublattice symmetry of graphene, and no band gap is opened at the Dirac point. But in the case of LiC$_{3x}$ (for instance, LiC$_6$), the situation is quite different, the C-C bonds in graphene sheet occupy two kinds of positions (called Kekulé construction)[16,28], which induces the bond symmetry breaking in graphene and produces a band gap at the Dirac point (a much more detailed discussion based on tight-binding model can be found in the article of Farjam *et al.*[28]). The mechanism of bond symmetry breaking should also take place in the cases of LiGe$_{3x}$, which enhances the effect of symmetry breaking and induce a larger band gap (NaGe$_{3x}$ and KGe$_{3x}$ also have larger gap, but not notably). In the case of AMGe$_2$/AMGe$_4$, the hybridization of the orbitals of germanium and AM atoms may play an important role to produce the smaller band gaps, as the right panel of Figure 4b-4c shows.

In order to investigate the mobility of charge carriers in germanene after adsorption, we calculate effective masses of electrons and holes from the dispersion relation of the $\pi^*$ and $\pi$ band near the Dirac point, using the definition

$$\frac{1}{m^*} = \frac{1}{\hbar^2}\frac{d^2 E}{d k^2} \qquad (3)$$

Since the effective masses of electrons and holes have almost same absolute value and are slightly anisotropic, we only give the average values of $m^*$ as the function of coverage in Figure 6. We find that $m^*$ tends to increase as $N$ increases, but becomes especially larger when $n = 3x$. This phenomenon is obviously related to the enhanced band gap due to the bond symmetry breaking mechanism. However, all the effective masses are small: the maximum is about 0.11 $m_e$ ($m_e$ is the mass of free electron), and the other values are less than 0.05 $m_e$. Hence, AM atoms adsorption will not affect charge carrier mobility in germanene obviously. This is important when we apply them in electronic devices.

All the investigated AM-covered germanene are not intrinsic semiconductor, since their band gaps are below the Fermi level. So they act as n-type channel in FET. When our materials are applied to FETs, a global band gap, not only a band gap at the Dirac point between the $\pi$ and $\pi^*$ band is needed. Unfortunately, the $\pi^*$ and $\pi$ bands of AMGe$_2$ and AMGe$_4$ are deformed seriously near the point $\Gamma$ (as shown in Figure 4b-4c), making the global band gap disappear, so AMGe$_2$ and AMGe$_4$ are not suitable to be applied in FETs. The maximum global band gap is 0.26 eV (LiGe$_6$), as shown in Figure 4d. The band gaps opened in germanene are in general smaller than those in silicene at the same coverage[16]. Hence, the on-off ratio of germanene-based FET with the surface adsorption will also be smaller, compared with the one based on silicene.

Since the SOC effects are much stronger in germanium atoms than those in silicon and carbon atoms, we also calculate the band structure of $LiGe_2$ and $NaGe_2$ with the inclusion of the SOC effects, as shown in Figure S2. The SOC effects cause a band gap of about 25 meV at the Dirac point in pure germanene, which is in good agreement with the previous work[26]. The band structures of $LiGe_2$ and $NaGe_2$ with and without the inclusion of the SOC effects are nearly indistinguishable, as shown in Figure S2b-S2c. The band gap at the Dirac point only increase by about 1 meV in $LiGe_2$/$NaGe_2$ after the inclusion of the SOC effects, this fact suggests that the SOC effects will only affect our results slightly.

## Conclusion

In summary, our *ab initio* calculations show that the adsorption of AM atoms on germanene is very strong, and it is able to open a band gap at the Dirac point in germanene, ranging from 0.02 to 0.31 eV. This band gap is mainly induced by the sublattice symmetry breaking mechanism, and the bond symmetry breaking mechanism also plays an important role in the case of $AMGe_{3x}$. The maximum global band gap we obtain is 0.26 eV, which is much larger than that obtained by applying an external electric field. The smaller effective masses near the Dirac point suggest the charge carrier mobility in germanene won't degrade upon AM atoms adsorption. All these results suggest the potential advantage of AM-covered germanene for high-speed effective FETs.

**Acknowledgements**

This work was supported by the National Natural Science Foundation of China (Nos. 11274016, 51072007, 91021017, 11047018, and 60890193), the National Basic Research Program of China (Nos. 2013CB932604 and 2012CB619304), Fundamental Research Funds for the Central Universities, National Foundation for Fostering Talents of Basic Science (No. J1030310/No. J1103205), and Program for New Century Excellent Talents in University of MOE of China.

**Supplementary Information**

Band structures of Li-covered germanene at the coverage of 1/8, 1/12, 1/16, 1/18, 1/24, and 1/32; band structures of germanene, LiGe$_2$ and NaGe$_2$ with and without the inclusion of the SOC effects.

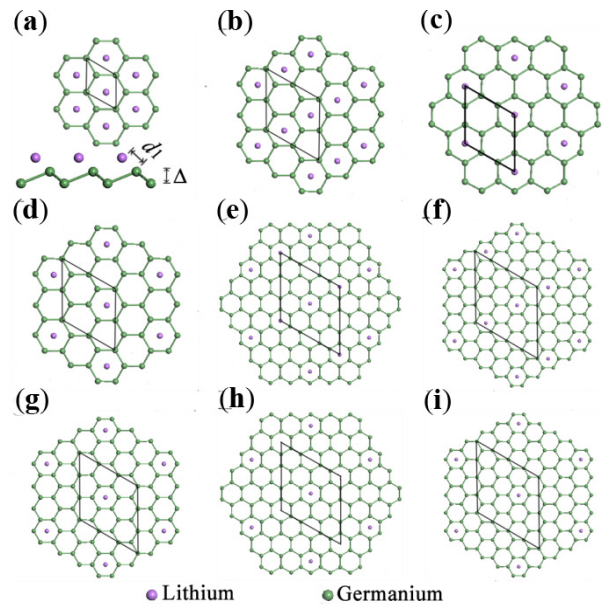

**Figure 1:** Structures of Li-covered Germanene. (**a**)-(**i**) The adsorption coverage is $N$ = 1/2, 1/4, 1/6, 1/8, 1/12, 1/16, 1/18, 1/24, and 1/32, respectively. The rhombi plotted in black line represent the supercells at different coverages.

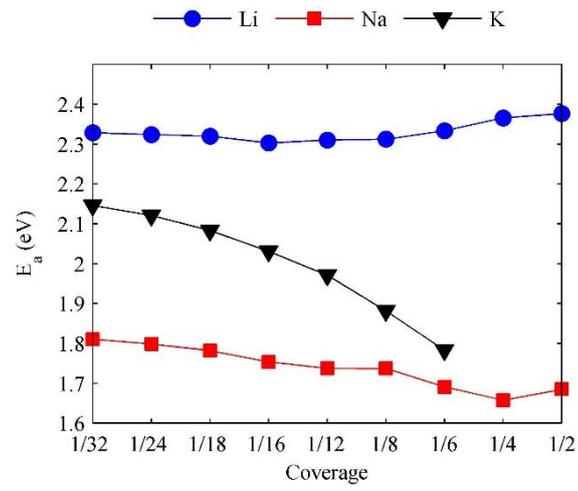

**Figure 2:** Adsorption energy of AM atoms on germanene per AM atom as a function of coverage.

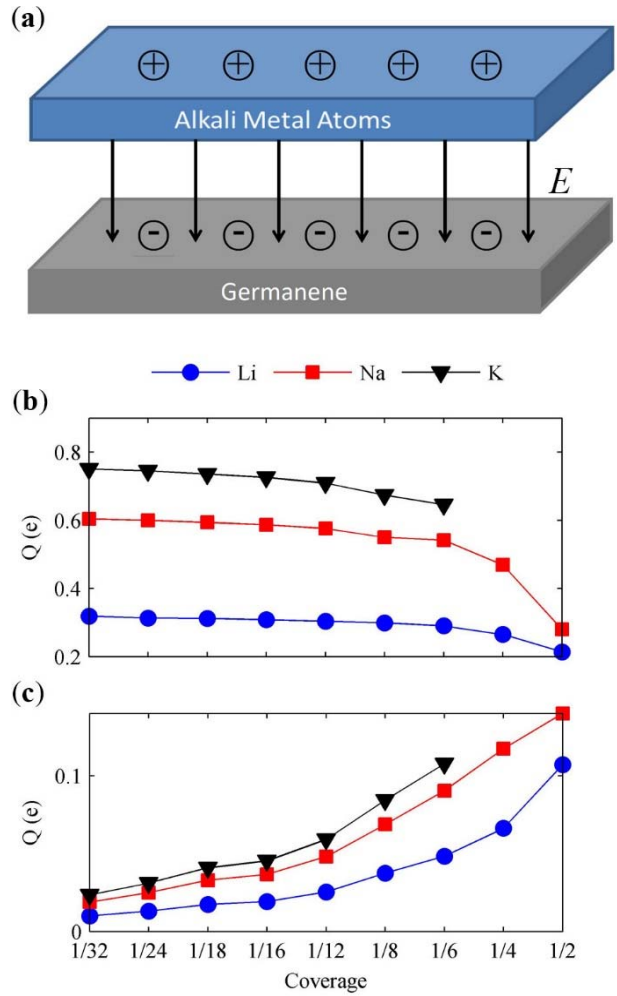

**Figure 3:** (**a**) Schematic of AM-covered germanene. *E* denotes the electric field which is given rise by transferred charge. (**b**)-(**c**) Transferred charge (using Mulliken population analysis) from per AM atom to germanene (**b**) per AM atom and (**c**) per germanium atom as a function of coverage.

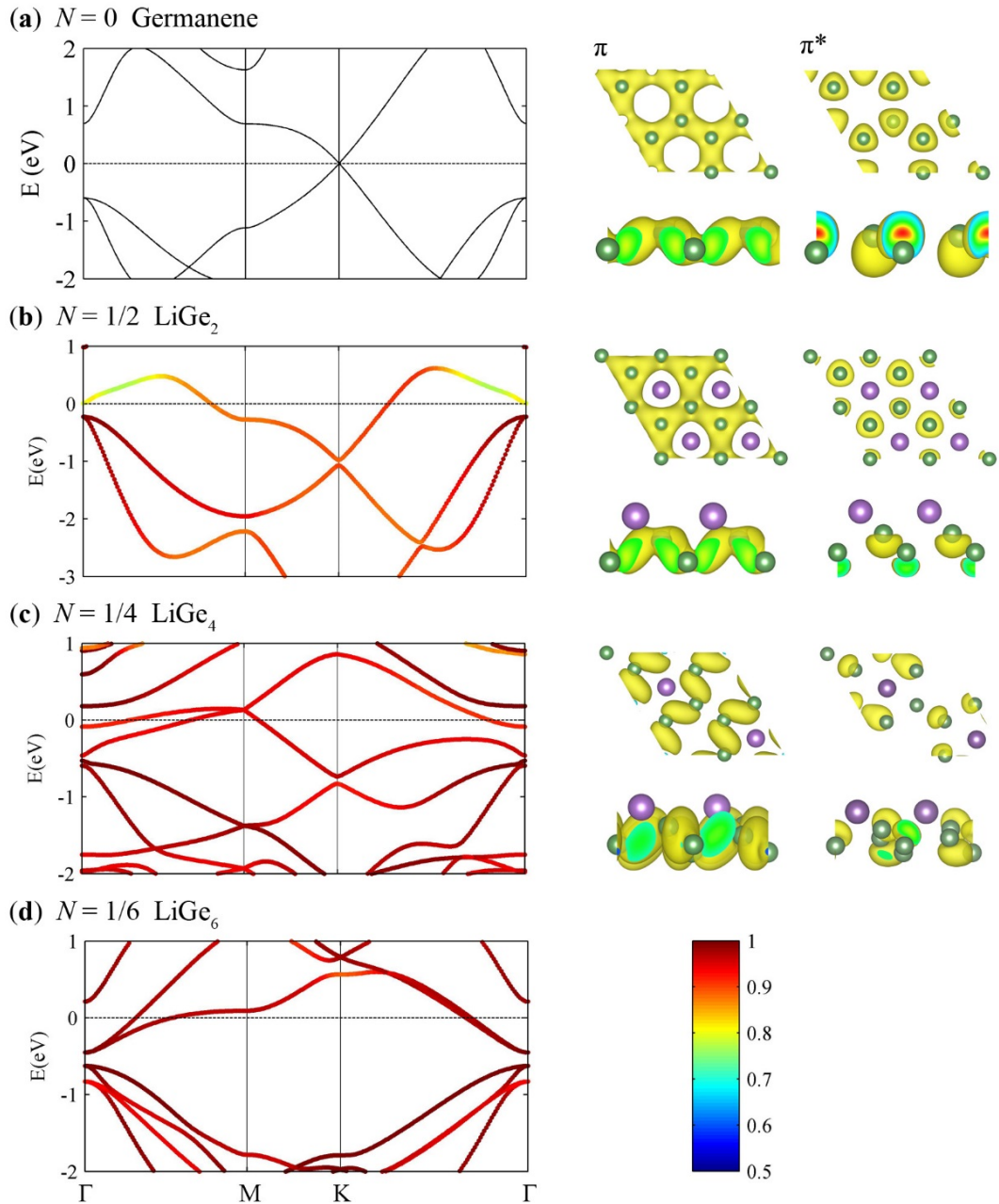

**Figure 4:** Left panel: Band structures of (**a**) pristine and (**b**)-(**d**) Li-covered germanene at coverage $N$ = 1/2, 1/4, and 1/6, respectively. The Fermi level is set to zero. Contribution from the germanium atoms are marked as different colors corresponding to the weight in (**b**)-(**d**), as the color map shows. Right panel: Isosurfaces of the orbitals of the π and π* band at the Dirac point of the correspond structures. All the images in the right panel were prepared in VESTA[29].

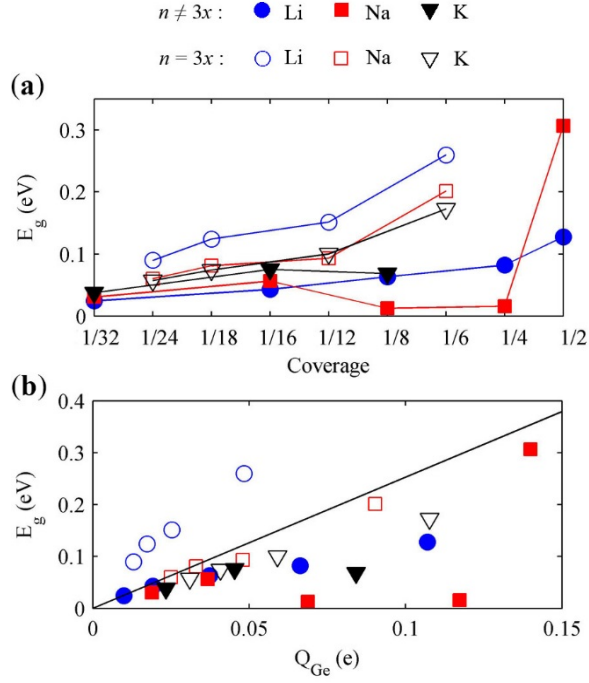

**Figure 5:** Band gap as a function of (**a**) coverage and (**b**) Mulliken transferred charge density $Q_{Ge}$ from AM atoms to germanene (per germanium atom). The solid line in (**b**) is the predicted band gap dependence on a given transferred charge density based on the calculated relationship between the band gap and electric field[15]. The electric field of a given transferred charge density is estimated from the parallel plate capacitor model.

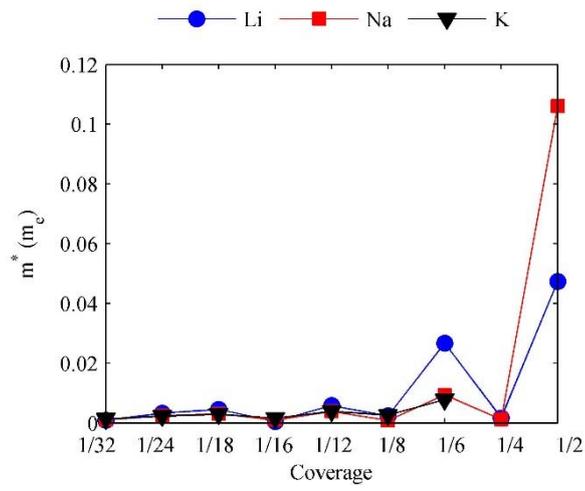

**Figure 6:** Effective mass of electron at the Dirac point as a function of coverage.